# Glass+Skin: An Empirical Evaluation of the Added Value of Finger Identification to Basic Single-Touch Interaction on Touch Screens

Quentin Roy[1,2], Yves Guiard[1], Gilles Bailly[1], Éric Lecolinet[1], Olivier Rioul[1]

[1] Telecom ParisTech, CNRS LTCI UMR 5141, Paris, France
[2] GE Healthcare, Buc, France
quentin@quentinroy.fr
{firstname}.{lastname}@telecom-paristech.fr

**Abstract.** The usability of small devices such as smartphones or interactive watches is often hampered by the limited size of command vocabularies. This paper is an attempt at better understanding how finger identification may help users invoke commands on touch screens, even without recourse to multi-touch input. We describe how finger identification can increase the size of input vocabularies under the constraint of limited real estate, and we discuss some visual cues to communicate this novel modality to novice users. We report a controlled experiment that evaluated, over a large range of input-vocabulary sizes, the efficiency of single-touch command selections with vs. without finger identification. We analyzed the data not only in terms of traditional time and error metrics, but also in terms of a throughput measure based on Shannon's theory, which we show offers a synthetic and parsimonious account of users' performance. The results show that the larger the input vocabulary needed by the designer, the more promising the identification of individual fingers.

## 1  Introduction

The number of buttons on small touchscreens (e.g. watches, wearable devices, smartphones) is strongly limited by the Fat Finger Problem [7,30,36]. Increasing the number of commands requires users to navigate through menus, lists or tabs, thus slowing down the interaction. This problem also arises on larger touch screens, such as tablets, where applications need to save as much space as possible for the display of objects of interest, rather than controls. For instance, users of photo-editing, 3D drawing, or medical imagery applications want to see large high-resolution images, but at the same time they want to see large command menus. Possible responses to this challenge are a drastic reduction in the number of available commands and functionalities (e.g., Photoshop offers 648 menu commands on a PC, and only 35 on a tablet [37]), and intensive recourse to hierarchical menus, at the cost of efficiency. For frequently used commands, the lack of hotkeys on touch-based devices badly aggravates this problem.

Many different approaches have been proposed in the literature to provide input methods that save screen real estate. Most of them rely on gestures [20,24,25,29,41,42] such as Marking menus [20,42], rolling gestures [32], multi-finger chords [23,37], finger-counting [3,4] etc. Another approach exploits additional sensors such as motion sensors or accelerometers [17] or pressure sensors [31]. In this paper we focus on finger identification and investigate to which extent it can augment the expressivity of touch input and allow larger command vocabularies while saving screen space.

Recognition of finger identity provides several advantages for command selection. Finger identification allows increasing the input vocabulary while being compatible with already existing interaction styles: For instance, the same button may serve to invoke different commands depending on which finger is pressing it. This strategy will increase the total number of commands for a *given interface*. But it will also reduce the number of necessary buttons for a given set of commands while maintaining a direct access to these commands (i.e. without the need to open menus, scrolling lists, etc.). Buttons can then be designed with larger sizes, thus easing interaction on small touchscreens. It is worth noticing that on such devices interaction is usually more constrained by (touch) input than by (visual) output. Because of the high pixel resolution of modern screens, icons — and often even text — can remain recognizable at sizes that preclude their selection using a finger tip. Finger identification can be exploited for displaying several icons (one for each available command) on *finger-dependent buttons* and thus make all commands discoverable, as we will see in section 3.

Finger identification can also serve to provide shortcuts for invoking frequent or favorite commands instead of opening context menus. For instance, "copy", "paste", "select" and other heavily used commands could be invoked in this way on smartphones.

Finger identification may facilitate the transition to complex chording gestures: Novice users will sequentially press two different buttons with two different fingers (e.g. index and middle fingers). More experienced users will execute these operations faster and faster until they perform these two actions simultaneously and perform a chording gesture.

To explore this promising modality a number of finger-identification prototypes have been described in the HCI literature, which in the near future are likely to become practical and robust.

Below we will call GLASS the usual input channel that considers only the *xy* coordinates of the contact on the screen, and GLASS+SKIN the augmentation of this channel with the skin (categorical, or non-metrical) coordinates, which requires finger identification.

In this paper, we try to better understand how interaction techniques relying on finger identification may help users invoke commands on touch screens. To progress towards this goal, we conducted a user study comparing the performance of finger-dependent buttons with traditional, finger-agnostic buttons, for various sizes of the command vocabulary. One of our concerns was to figure out when finger-identification starts outperforming traditional button-based interfaces.

The results showed that if the standard channel is perfect for very few commands, it is soon outperformed by the GLASS+SKIN option, in a given amount of real estate, as the number of commands increases. The main finding is that with GLASS+SKIN the error rate increases at a considerably reduced pace with vocabulary size, which makes it possible to handle much larger sets of commands. We found that the maximum obtainable bandwidth (or, more precisely, the maximal level of possible throughput, in Shannon's [33] sense) is higher and that users can handle larger vocabularies with finger-sensitive than finger-agnostic touch detections.

## 2      Related Work

### 2.1      Augmenting the Expressivity of Touch Input

In the face of the small size of the screen and the fat finger problem [19, 30,36], several modalities have been proposed to augment the expressivity of touch input. The most widespread of all seems to be multi-touch input [22], especially with the most successful zoom-and-rotate gesture that the iPhone popularized. One particular exploitation of the multi-touch was Finger-Count [3,4], which determines command selection based on just the *number* of finger contacts from both hands.

Other modalities have also been proposed such as touch in motion [17] or pressure+touch input [7,15,28,31], whose input bandwidth unfortunately is low because selection time is long (from ~1.5s to more than 2s with no feedback) and whose users distinguish hardly more than 5-7 values [28].

Our motivation is to understand what happens if screen and skin coordinates of touch input are distinguished. In this spirit, Roudaut et al. recognize the signature of fingers' micro-rolls on the surface [32]. Wang et al. used the orientation of the finger to control parameters [39]. Holz and Baudish detect fingerprint to improve touch accuracy [19]. And more recently, TapSense uses acoustic signatures to distinguish the taps from four different part of users' fingers: tip, nail, knuckle and pad [14].

In this class of interaction, proper finger identification — with screen and skin coordinates jointly taken into account — seems highly promising [1,6,9,11,12,18,23,26,37,40]. Many studies have concentrated on triggering finger-dependent commands or action [1,6,23,26,37]. For instance, Adoiraccourcix maps different modifiers to the fingers of the non-dominant hand and different commands to fingers of the dominant hand [12]. Finger identification can also be coupled with physical buttons as in recent Apple Smartphones [35,40]. The advantage of this method is that the identification can be performed even if the button is not pressed, adding a supplementary state to the interaction [8]. Finger-dependent variants of chords and Marking Menus have also been investigated [23].

Some researchers have examined the discoverability of finger-dependent commands. For example, Sugiura and Koseki [35] identify the finger as soon as a user touches a (physical) button. They use this property to show a feedback on the corresponding command name prior to the actual button press. This, however, is not compatible with most touch systems, which more often than not lack a passive state [8]. In Au et al. [1] a menu is displayed showing the commands under each finger, but users must depress their whole hand on the surface to invoke it. In section 3.3 we will consider various techniques of informing users about the availability of finger-dependent commands.

### 2.2 Finger Identification Technologies

Under certain circumstances, fingers can be identified using the default hardware of hand-held computers. Specific chord gestures are typically used for this purpose. Assuming a relaxed hand posture, the user must touch the surface with a certain combination of fingers or perform a specific temporal sequence [1,37]. Some other multi-touch techniques such as MTM [2] or Arpege [10] do not directly identify fingers but infer them based on the likely positions of individual fingers relative to some location of reference.

Computer-vision can be used to identify fingers without requiring chording gestures. The camera can either be located behind the interactive surface such as with FTIR multi-touch tables (e.g. [23]) or placed above with a downward orientation (e.g. [5]). The idea is to compare fingertip locations (obtained through computer-vision) with touch event locations (provided by the interactive surface). Basic solutions identify fingers by considering their relative positions. But this approach fails if some fingers are flexed (e.g. [9]). Markers can be attached to the fingers to solve this problem (e.g. color [40] or fiduciary tag [26]). But, this cumbersome solution, which demands that the users be instrumented, is workable only in research laboratories. Some commercial systems are able to track the mid-air motion of individual fingers (e.g. Microsoft Kinect and Leap Motion). This approach makes it possible to identify which fingers come in contact with a surface [21].

Hardware-based approaches have also been proposed. Sugiura and Koseki [35] used a fingerprint scanner to identify fingers. They were able to trigger finger-dependent commands but not to track finger positions. Holtz and Baudish extended this work to touchpads [19] and more recently to the touch-screen of interactive tables [18]. Another approach consists of analyzing EMG signals on the forearm to determine which finger is applying pressure to the surface [6]. In yet another approach, Goguet et al. attached *GameTraks*[1] to user's fingers [11,12]. Of course, digital gloves can also serve to track user fingers [34]. A drawback of these approaches is that they require user instrumenting and/or a calibration phase.

---

[1] *GameTrak* is a game controller designed for the Sony PlayStation 2. It is equipped with two retractable strings usually attached to the player's wrists. It is able to track the 3D position of the attached limbs on top of the device.

## 3  GLASS+SKIN: A Class of Promising Interaction Techniques

Several widgets such as toolbars or menus exclusively rely on the spatial arrangement of buttons on the screen. During interaction with these widgets the system only exploits the *screen* coordinates of finger contacts to interpret the decisions of users. In this section, we show how finger identification can offer interesting properties to improve command selection on touch screens. In this section, we give some insights in how application designers may leverage GLASS+SKIN, a class of interaction techniques that augment traditional interaction with finger identification.

### 3.1  Multi-Function Buttons

**Increasing the input vocabulary.** With GLASS+SKIN input, a button can invoke more than one command. From the moment individual fingers are identified, more commands can be handled for the same amount of screen real estate. For instance, the main screen of the iPhone can provide a direct access to 20-24 applications (a 4x5 or 4x6 array of buttons, depending on the model). Whether useful or not, with five fingers discriminated, these numbers could be multiplied by 5.

**Reducing the number of buttons.** More interestingly, perhaps, on a given screen with a given set of commands, GLASS+SKIN input can just as well reduce the number of buttons. Direct access to these commands is maintained, without the need to open a hierarchical menu or scroll a list. Moreover, if more space is available, buttons can be designed with larger sizes, facilitating the interaction with small touchscreens.

**Compatibility.** One concern is to make GLASS+SKIN interaction compatible with users' habits. To this end the default button behavior might be assigned to the index finger that most users prefer for touch-screen interaction. Only experienced users would be concerned with the set of additional commands (four extra possibilities per button).

**Input vs. Output.** If a button can invoke different commands, it should communicate the different options it offers. It is worth noticing that interaction is usually more constrained by (touch) input than by (visual) output on such devices. Because of the high pixel resolution of modern screens, icons - and even text to a certain extent - can remain recognizable at sizes for which they could hardly be selected using a finger. Displaying several icons (one for each available command) on multi-function buttons it is thus possible to make all commands discoverable. After all, buttons on hardware keyboard already contain several symbols that can be accessed from different modifiers (i.e. Ctrl, Shift, Alt).

**Cancel.** Users pressing a button with the wrong finger can cancel the current selection by moving their finger away from the target or just waiting for a delay. The mapping then appears and users can release the finger without triggering a command.

### 3.2 Menus

GLASS+SKIN can reduce the needs for menus from small to medium applications. However, when the number of commands is very large, it is difficult to avoid menus, which are useful for organizing commands. This section considers how GLASS+SKIN fares with menus.

Menu shortcuts, such as keyboard shortcuts, are generally not present on mobile devices. We propose to use finger identification as a substitute for menu shortcuts on touchscreens. This makes it possible both to interact in the usual way (by opening menus and clicking on their items) and to activate frequent or favorite commands quickly (by pressing the appropriate finger on the touchscreen). Finger identification can thus serve to (partly) compensate for the lack of keyboard shortcuts on mobile devices (see Figure 2c).

**Context menus.** GLASS+SKIN can provide an expert mode to context menus. Novice users continue to press and wait for a delay to open the menu. However, more experienced users can invoke commands without waiting for the delay. The five most frequent or favorite commands of the menu are assigned to the five fingers. This can be especially useful for selecting repeatedly used commands such as "copy", "paste" or "select". Alternatively, one can choose to sacrifice one shortcut to remove the menu delay: e.g., the thumb could open the menu instantly.

**Menu bar, Tool bar and Folders.** Some persistent buttons give access to pull-down menus. In this case, the index finger is still used to navigate in the hierarchy of commands as usual. However, the other fingers provide a direct access (shortcuts) to favorite (or frequent) menu items deeper in the hierarchy. Suppose the index finger is still used to open a folder on smartphone. The four remaining fingers are shortcuts to select pre-defined items within this folder. This class of interaction strongly differs from approaches relying on finger chords [4,10,37] which specify not one but several contact points (one per finger) making it difficult to predict their behavior on small widgets (smaller than the required surface to contain all contact points).

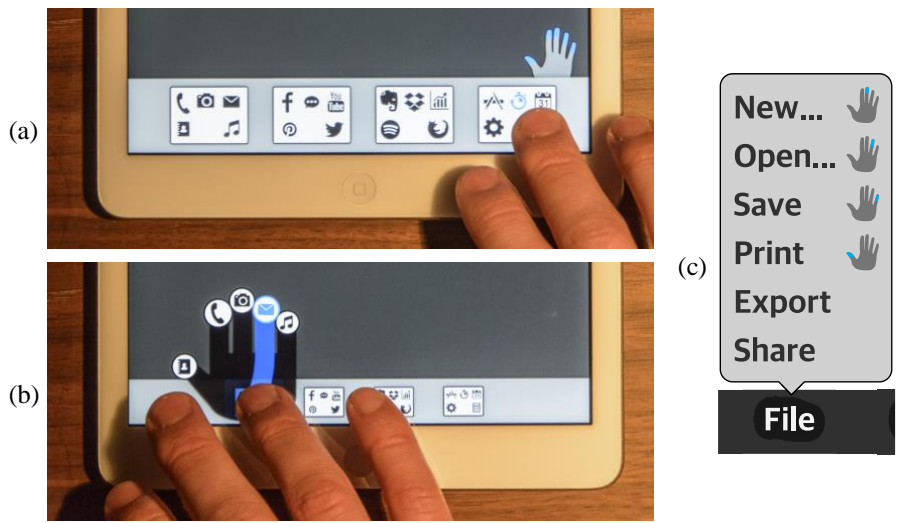

**Figure 2.** GLASS+SKIN menu instances.

### 3.3 Communicating GLASS+SKIN

**Discovering.** Some users can be unaware of this novel input modality. Some visual cues can help them to discover this modality without using video tutorial or documentation. We consider two of them in this project illustrated in Figure 2a and Figure 2b. The first one is static and displays a ghost hand on top of the toolbar to indicate that different fingers can be used. The second one is dynamic and shows a short animation showing several surface contacts with different fingers. Further studies are necessary to evaluate the ability of users to understand the meaning of these icons.

**Mapping.** When a button has several commands, it is important to communicate which finger activates which command. Figure 2 illustrates 3 visual cues to understand the mapping. The first one uses the location of the icon inside the button to convey the target finger. The second one builds on the previous and appears only on demand. Users should press and wait for 100ms to see the mapping. This approach reduces the total amount of information on the screen for expert users but can be less intuitive for novice users. The last example uses fingers as a menu shortcut. Symbols representing the target finger are shown on the right of the command name similarly to keyboard shortcuts on linear menus.

**Toward chording gestures.** Finger identification may facilitate the transition to complex chording gestures: Novice users will sequentially press two different buttons with two different fingers (e.g. index and middle fingers). More experienced users will execute these operations faster and faster until they perform these two actions simultaneously and perform a chording gesture.

### 3.4 Limitations

GLASS+SKIN also has some limitations. For instance, the different interaction techniques are not compatible with each other, e.g. a GLASS+SKIN button cannot launch five applications and open a menu. Designers should make compromises according to the users' needs and the coherence between applications/systems.

In some situations, it can be difficult to use a specific finger on the touch screen. Though the current smartphone trend is to large screens precluding a single hand use, some users still often use their smartphone this way. In this case, not only is the novel input resource unavailable to users, but errors may also arise if the application does not consider the thumb as the default finger. One solution would consist of constraining GLASS+SKIN to a subset of applications (e.g. games). Another would require sensing and recognizing grab [13] to avoid accidental activations. GLASS+SKIN is probably more useful for tablets or watches where "thumb interaction" is less common.

## 4   A Controlled Experiment

The experiment was designed in light of Shannon's theory [33]. A communication channel permits a source of information (the user) to transmit information to a destination (the system), the user's hand serving as the emitter and the touch screen as the receiver of the coded message. The code shared by the source and the destination is some mapping of a set of touch events to a set of commands. The larger the sets, the more entropy in our vocabulary of commands. For simplicity, below we will assume equally probable commands: in this case the input entropy (or the vocabulary entropy $H_V$) is just the $\log_2$ of the number of possible commands.

Although we will not ignore traditional time and error metrics, our analysis will focus on the *throughput* (*TP*), the rate of successful message transmission over a communication channel. We simply define the *TP* (in bits/s) as the ratio of Shannon's mutual information transmitted per command to the time taken on average to enter the command. Our main concern is the particular vocabulary size that maximizes the throughput — i.e., the optimal level of vocabulary entropy ($H_{opt}$, in bits) — in the two conditions of interest. In the GLASS condition, our baseline, the command vocabulary leveraged only the entropy offered at the surface of the glass (the $\log_2$ of the number $N$ of graphical buttons), as usual; in the GLASS+SKIN condition we also leveraged the entropy available on the skin side. The vocabulary size is then $NN'$, where $N'$ denotes the number of identifiable bodily regions that may touch the screen (in practice the experiment involved the five finger tips of the right hand). The entropies of these two independent variables add up — i.e., $\log_2(NN') = \log_2(N)+\log_2(N')$ — allowing the creation of larger command vocabularies. Our problem was to experimentally evaluate the actual usability of such enlarged vocabularies.

We were able to formulate several straightforward predictions.
(1) As the vocabulary entropy is raised, the amount of transmitted information $I_t$ must level off at some point, just as has long been known to be the case in absolute-judgment tasks [27].
(2) On the other hand, mean selection time $\mu_T$ must increase about linearly with $H_V$, due to Hick's law and Fitts' law.
(3) It follows from (1) and (2) that the dependency of $TP = I_t/\mu_T$ upon $H_V$ must be bell shaped — for any given input technique there must exist an optimal level of entropy.

Thus we will focus on the maximum of *TP* ($TP_{max}$) reached at the optimal level of entropy, and on the particular level of entropy, which we will designate as optimal ($H_{opt}$), at which that maximum takes place. One faces two independent pieces of empirical information: The higher the $TP_{max}$, the better the information transmission; the higher the $H_{opt}$, the larger the range of usable vocabulary sizes.

We conjectured that when contacting a touch screen users have control not only over the selection of one screen region, but also over the selection of one region of their own body surface. Put differently, the glass surface and the skin surface should be usable as more or less independent input channels. Therefore both $TP_{max}$ and $H_{opt}$ should be raised with GLASS+SKIN, relative to the GLASS baseline.

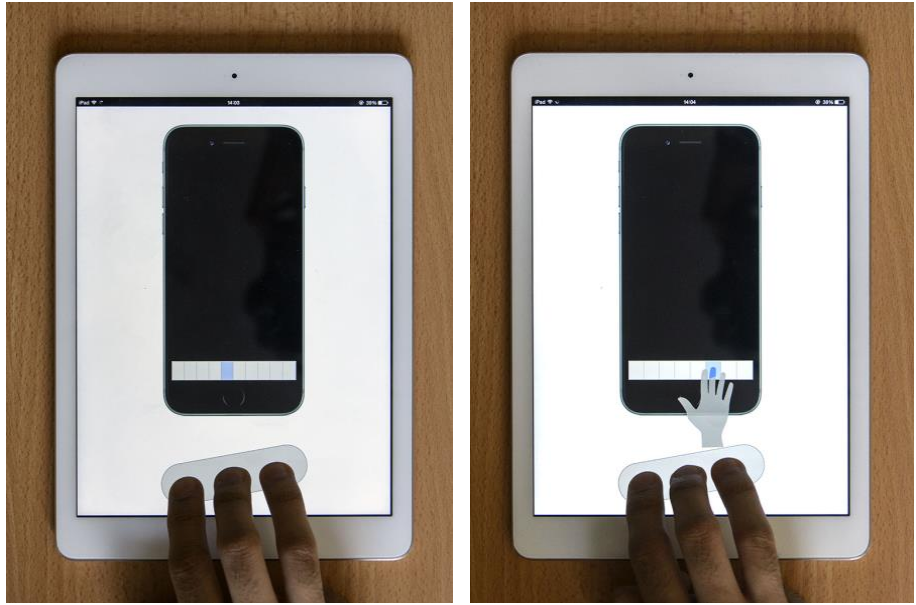

**Figure 3.** The display at the time of appearance of the stimulus in the GLASS (left) and the GLASS+SKIN (right) conditions.

### 4.1 Participants and apparatus

14 right-handers (5 females) ranging in age from 21 to 33 years, recruited from within the university community in our institution, volunteered.

The apparatus consisted of an iPad tablet (9.7 inches / 24.6cm in diagonal) reproducing the screen of an iPhone (see Figure 3). A start button, on which participants had to rest their forefinger, middle finger, and ring finger, was displayed below the smartphone, so as to standardize the position and the posture of the hand at trial start. The target area was displayed as a horizontal layout extending over the complete width of the phone screen (2.3 inches / 59 mm), simulating the common toolbars/docks of smartphones. Buttons height was a constant 0.90mm (as on an iPhone). We considered manipulating the size of the target area and the layout of buttons as factor, however we decided to focus on this configuration to keep the experiment short enough. Pilot studies, in which we also tested 2D grid layouts, showed that simple 1D layouts produced essentially the same results. The software was implemented with Javascript.

Table 1. Number of commands, number of buttons, and horizontal button size

| Number of Commands | Number of Buttons GLASS | Number of Buttons GLASS+SKIN | Button Width GLASS | Button Width GLASS+SKIN |
|---|---|---|---|---|
| 5  | 5  | 1  | 12mm / 0.46in    | 58mm / 2.3in   |
| 10 | 10 | 2  | 5.8mm / 0.23in   | 29mm / 1.2in   |
| 15 | 15 |    | 3.9mm / 0.15in   |                |
| 20 | 20 | 4  | 2.9mm / 0.11in   | 15mm / 0.58in  |
| 30 | 30 | 6  | 1.9mm / 0.077in  | 9.7mm / 0.38in |
| 40 | 40 | 8  | 1.4mm / 0.058in  | 7.3mm / 0.29in |
| 50 |    | 10 |                  | 5.8mm / 0.23in |
| 70 |    | 14 |                  | 4.2mm / 0.16in |

## 4.2   Method

**Task and Stimulus.** In response to a visual stimulus, participants were to select a command as fast and accurately as possible by touching a target button highlighted in gray. In the GLASS+SKIN condition, a ghost hand was also shown (Figure 3 right), the target finger, highlighted in blue, coinciding with the target button.

**Procedure.** The participants started the trial by placing their three longer fingers on an oblique start button located at the bottom of the screen. The system responded by presenting the stimulus (depending on the condition either just a button highlight or a button highlight plus the ghost hand). The stimulus remained as long as the start button was occupied.

If correctly hit the target button turned green. A mistakenly-hit button was highlighted in red. If for any reason no touch was recorded, the participant was supposed to return to the start button to reset the trial. The finger identity of touch events was not recorded. Video recordings in a pilot experiment using our ghost-hand stimuli having revealed a remarkably low error rate for finger selection (2.3% on average, $\sigma = 2.0\%$), it seemed reasonably safe to trust participants. Video recordings of a sample of 3 participants during the present experiment showed similar results (1.5% on average, $\sigma = 0.52\%$).

We used a within-participant design. The order of techniques and the size of the command vocabulary were counter-balanced between participants with Latin squares, each command randomly appearing three times per block. The total duration of the experiment was about 30min/participant. Overall the experiment involved 14 participants x (5+10+15+20+30+40+5+10+20+30+40+50+70) trials x 3 iterations of each trial type = 14,490 selection movements.

**Vocabulary Size**. Relying on pilot data, we chose to use 5, 10, 15, 20, 30, and 40 possibilities for GLASS and 5, 10, 20, 30, 40, 50 and 70 possibilities for GLASS+SKIN. The more possibilities in a 60mm-wide array, the smaller the target. With GLASS+SKIN, the number of *screen* targets was divided by 5 (Table 1).

### 4.3 Results

**Classic Time/Error Analysis**. The relevant dependent variables are the reaction time (*RT*, the time elapsed between stimulus onset time and the release of the start button), movement time (*MT*, the time elapsed between release of the start button and the first detection of a screen contact), and the error rate.

Significance was estimated using ANOVA. Non-common values of *number of commands* are ignored in the time and error analysis so that the comparisons between GLASS and GLASS+SKIN are relevant.

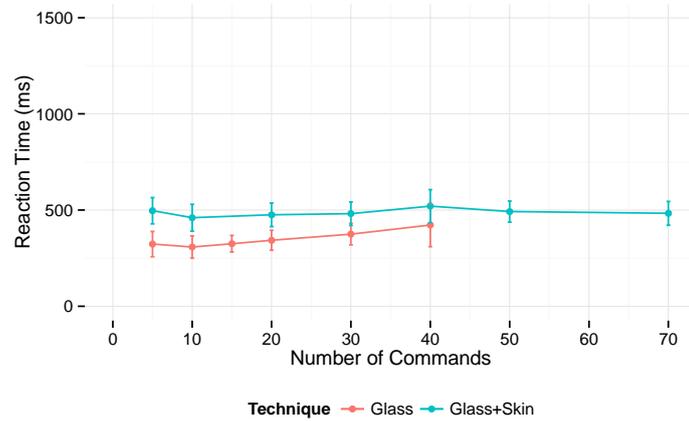

**Figure 4.** Mean *RT* vs. the number of commands for each condition.

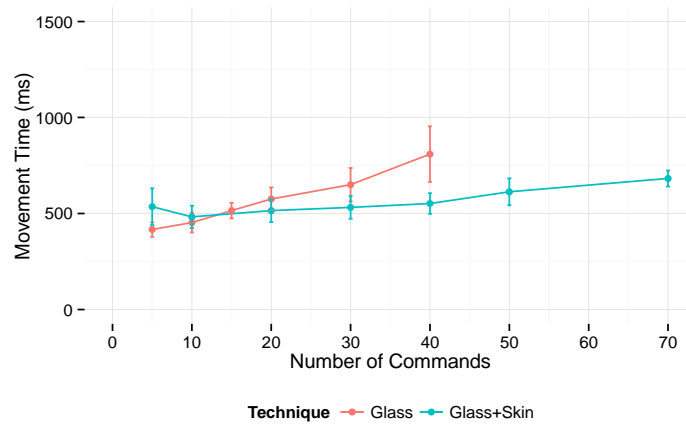

**Figure 5**. Mean *MT* vs. the number of commands for each condition.

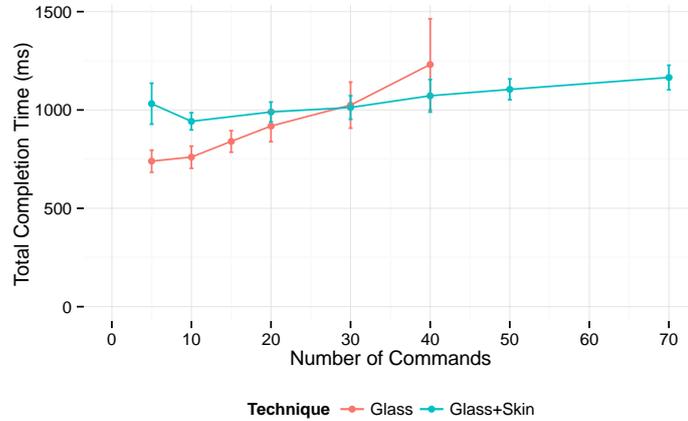

**Figure 6**. *Total Time* vs. the number of commands for each condition.

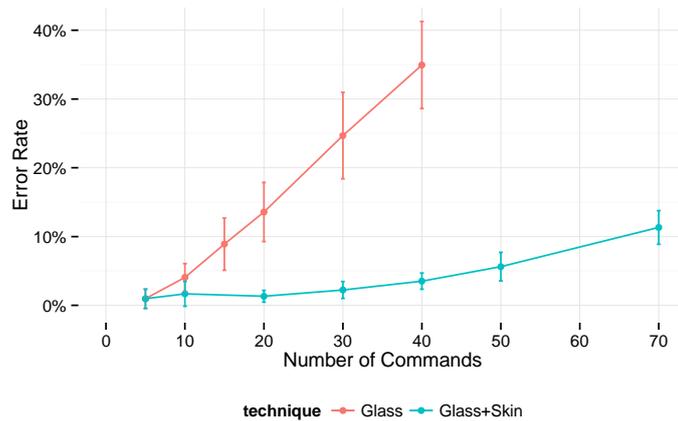

**Figure 7**. *Error rate* vs. the number of commands for each condition.

Reaction Time (*RT*) was faster with GLASS than GLASS+SKIN (Figure 4), a result observed in all our 14 participants (*p*<.001). The mean difference, computed over the common range of abscissas, was 132ms. The number of commands slightly affected *RT* for GLASS, but not GLASS+SKIN.

Overall, mean movement time (*MT*) was shorter with GLASS+SKIN than GLASS (Figure 5). The mean difference amounted to 43ms. The effect of vocabulary size, more pronounced on *MT* than *RT*, was approximately linear, with a steeper slope for GLASS *($F_{1,13}$=26, p=.0001)*.

On average, over the common range of abscissas, total task completion time (*TT=RT+MT*) was slightly (89ms) higher with GLASS+SKIN (Figure 6). Much more importantly, *TT* increased at a much slower pace as vocabulary size was raised ($F_{1,13}$=16.5, *p*=.001). With more than 30 commands, GLASS+SKIN was faster.

We conclude from this classic analysis of our data that taking into account the skin (categorical) coordinates of the touch event *together with* the glass (metrical) coordinates of the event enhances both the speed and accuracy of input selection, for large vocabularies. The error rate increasing at a considerably reduced pace with vocabulary size, GLASS+SKIN makes it possible to handle much larger sets of commands (figure 7). This error rate does not include potential mistakenly-used finger. However, video recordings from a sample of 3 participants showed that it is particularly rare (1.5% of the trials on average, σ = 0.52%).

**Information-Theoretic Analysis**. One reason why we felt the throughput (*TP*) analysis was worth a try is because this quantity combines the speed and the accuracy information into a single, theoretically well-justified quantity. Let us ask how the amount of successfully transmitted information $I_t$ (bits), and then the *TP* (bits/s) vary with the entropy of the vocabulary (simplified to $\log_2 N$ and $\log_2 NN'$).

In both conditions, $I_t$ tended to level off as $H_V$ was gradually raised, confirming the limited capacity (in bits per selection) of the tested transmission channels. Had we investigated larger vocabularies, the leveling off would have been more spectacular, but exploring very high levels of entropy is not just time consuming — also recall that in general humans hate to make errors. Below we will report evidence that in fact our range of *x* values, chosen in light of our pilot results, was adequate.

The two curves of Figure 8 tend to asymptote to different capacity limits. With GLASS+SKIN not only was the average amount of transmitted information higher than it was with GLASS (this difference was observed in all 14 participants), the capacity limit suggested to the eye by the curvature of the plot was invariably higher (14/14).

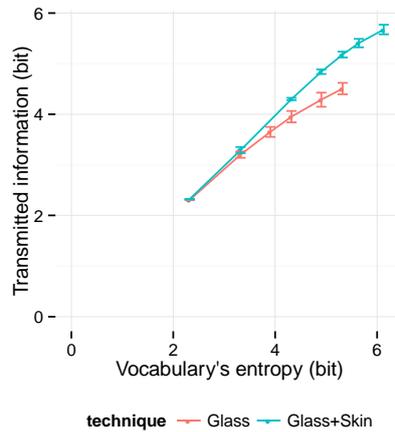

**Figure 8.** $I_t$ vs. $H_V$, for each condition

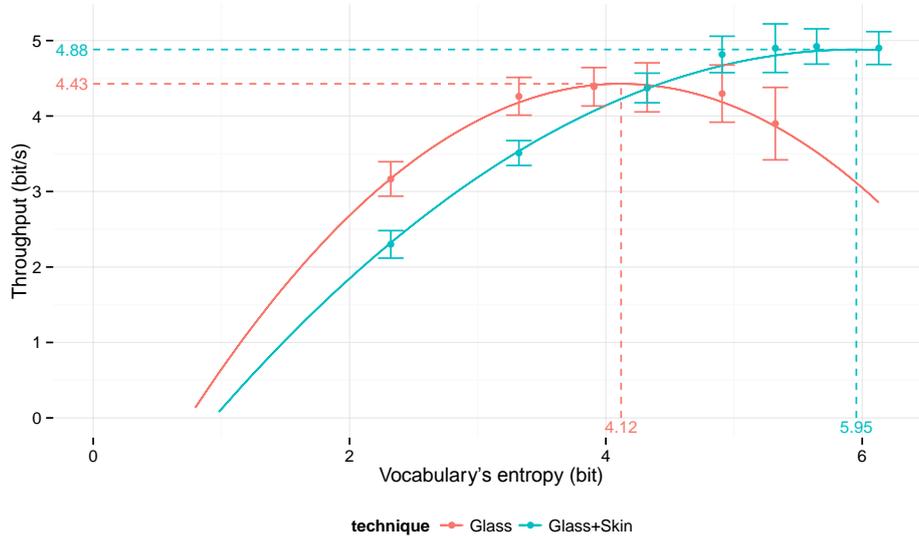

**Figure 9**. *TP* vs. $H_V$, for each condition.

We may now turn to the *TP*, which in both conditions reached a maximum, as predicted (Figure 9). Fitting second-order polynomials to the data, we obtained:

$y = -0.389x^2 + 3.2043x - 2.1714$ ($r^2 = .985$) for GLASS and

$y = -0.1941x^2 + 2.3115x - 2.0004$ ($r^2 = .997$) for GLASS+SKIN.

From these equations, shown graphically in Figure 9, one can estimate the *xy* coordinates of the maxima (both maxima take place within the tested range of entropies, and so no extrapolation is required):

$TP_{max}$ = 4.43 bits/s at an entropy level of 4.12 bits for GLASS and

$TP_{max}$ = 4.88 bits/s at an entropy level of 5.95 bits for GLASS+SKIN.

Thus a single figure illustrating the *TP* suffices to show unambiguously that the GLASS+SKIN resource entails two independent improvements. One is a 10.1% increase of the *TP*, meaning a more efficient transmission of information from the user to the system. The other is a 44.4% increase of optimal input entropy, meaning that much larger sets of commands can be effectively handled.

**Differential Finger Performance**. Obviously our fingers are not all equally suitable to serve in the GLASS+SKIN approach, for reasons unlikely to have much to do with entropy. Figure 10 suggests, unsurprisingly, that our best performer is the forefinger and the worse is the pinky, as summarized most compactly by the *TP* data of Figure 10 f. Any attempt to leverage the GLASS+SKIN principle in some interaction technique should probably consider focusing on the three central fingers of the human hand. Bearing in mind the current proliferation of small devices, however, the possibility to multiply the vocabulary by just 3 (thus adding up to $\log_2 3 = 1.58$ bits to $H_V$) seems of non-negligible interest.

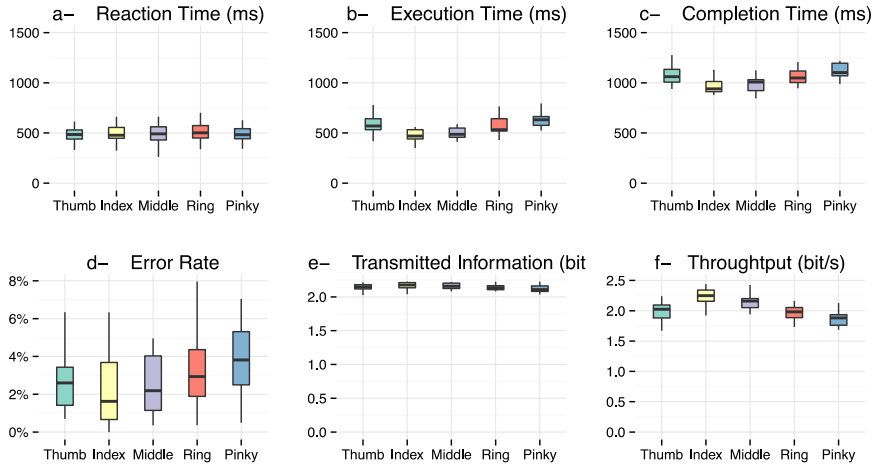

**Figure 10**. Multiple quantitative characterization of finger performance

## 5 Conclusion and perspectives

In view of the specialized literature there is no doubt that finger identification has a potential to considerably enhance input expressivity on touch screens in the near future, even (but not exclusively) in the simplest case of single-touch input that was considered in this research. The data of the above-reported experiment suggest that touch-screen input may certainly benefit from the substantial functional parallelism of the skin and glass channels, as we called them. We discovered that surprisingly little effort is demanded of users to adapt their hand posture, during hand motion to the target object, so as to touch this one target with this one finger. Importantly, our pilot experiments revealed that the latter choice, unlike the former, is essentially errorless.

One reason why the skin channel is of interest in the face of the real-estate scarcity challenge is that exploiting this additional channel makes it possible to increase the width of hierarchical command systems and hence to reduce their depth. For example, with just three fingers rather than one, and the GLASS+SKIN principle, one may escape the problematic design imagined by Apple in which 20 control buttons are displayed on a watch (Apple Watch Sport).

In the theoretical introduction to our experiment we offered a schematic view of the input problem. In particular, we left aside the complex *code* issue (movement-to-command mapping) and we deliberately ignored the fact that in the real world some commands are far more frequent than others, meaning the real levels of entropy are less than we assumed. These obviously are subtle and important issues that will deserve sustained attention in future research if the GLASS+SKIN principle is ever to be optimally leveraged. One important question in this direction is, What part of the information should be transmitted through which channel? One obvious constraint is that while screen regions (buttons) can be, and are invariably marked with text or symbols reminding users of which button does what, it is more difficult to imagine tricks that will remind which fingers does what without consuming screen space.